\documentclass[pra,
                twocolumn,
                preprintnumbers,
                amsmath,
                amssymb,
                showpacs,
                superscriptaddress]{revtex4-1}
\usepackage{graphicx} % Include figure files
\usepackage{dcolumn}  % Align table columns on decimal point
\usepackage{bm}       % bold math
\usepackage{amsfonts}
\usepackage{mathtools}
\usepackage[colorlinks=true,linkcolor=blue,citecolor=red,urlcolor=blue]{hyperref}
\usepackage{color}
\usepackage{natbib}
\newcommand{\eq}[1]{Eq.~(\ref{#1})}
\newcommand{\ful}{\mbox{C$_{60}$}}
\newcommand{\fulp}{\mbox{${\mbox{C}_{60}}^+$}}
\newcommand{\fuls}{\mbox{\scriptsize C$_{60}$}}

\begin{document}

\title{Photoemission from hybrid states of Cl@$\ful$ before and after a stabilizing charge transfer}

\author{Dakota Shields}
\affiliation{%
Department of Natural Sciences, D.L.\ Hubbard Center for Innovation,
Northwest Missouri State University, Maryville, Missouri 64468, USA}

\author{Ruma De}
\email[]{ruma@nwmissouri.edu}
\affiliation{%
Department of Natural Sciences, D.L.\ Hubbard Center for Innovation,
Northwest Missouri State University, Maryville, Missouri 64468, USA}

\author{Mohamed El-Amine Madjet}
\affiliation{%
Qatar Environment and Energy Research Institute, Hamad Bin Khalifa University, P.O. Box 34110, Doha, Qatar}

\author{Steven T. Manson}
\affiliation{%
Department of Physics and Astronomy, Georgia State University, Atlanta, Georgia 30303, USA}

\author{Himadri S. Chakraborty}
\email[]{himadri@nwmissouri.edu}
\affiliation{%
Department of Natural Sciences, D.L.\ Hubbard Center for Innovation,
Northwest Missouri State University, Maryville, Missouri 64468, USA}

\date{\today}

%\pacs{61.48.-c, 33.80.Eh, 36.40.Cg}

%%%%%%%%%%%%%%%%% END OF PREAMBLE %%%%%%%%%%%%%%%%

\begin{abstract}
Photoionization calculations of the endofullerene molecule Cl@$\ful$ with an open-shell chlorine atom are performed in the time-dependent local density approximation (TDLDA) based on a spherical jellium model. Cross sections for atom-fullerene hybrid photoemission studied show the effects of the hybridization symmetry, the giant plasmon and the molecular cavity. Comparisons with the results of Ar@$\ful$ provide insights in the role of a shell-closing electron and its influence on the dynamics. The results for Cl@$\ful$ are further compared with those of a more stable, lower energy configuration that results after a $\ful$ electron transfers to Cl forming Cl$^-$@$\fulp$. This comparison reveals noticeable differences in the ionization properties of the antibonding hybrid state while the bonding hybrid remains nearly unaltered showing a magnification covering the entire giant plasmon energy range. 
\end{abstract}

\maketitle 

\section{Introduction}
Significant success in the synthesis of endofullerene molecules -- systems of an atom or a smaller molecule incarcerated within the fullerene cage~\cite{popov2013} -- has spawned a series of experiments~\cite{mueller2008,kilcoyne2010,phaneuf2013} using merged beam techniques at Berkeley Advanced Light Source. Experiments accessed photoionization properties of these materials in gas phase. In addition, endofullerenes, being natural entrapment of atoms, have led to a series of theoretical studies of the effects of ionizing radiation on these systems, i.e., photoionization~\cite{dolmatove2009,chakraborty2015}. Such fundamental spectroscopic knowledge is particularly useful due to a broad horizon of applied importance of these materials in, namely, (i) solid state quantum computations~\cite{harneit2007,ju2011}, (ii) improving the superconducting ability of materials~\cite{takeda2006}, (iii) biomedical fields~\cite{melanko2009}, (iv) contrast-enhancement research for magnetic resonance imaging (MRI), (v) improving organic photovoltaic devices~\cite{ross2009}, and even in (vi) astrophysics~\cite{becker2000}. 

Endofullerenes confining open-shell atoms have potential applied interests of rather exotic nature~\cite{lawler2017}. For instance, N@$\ful$ has attracted interest due to its uniquely long spin relaxation times driven by the confinement~\cite{morton2007}. Electron paramagnetic resonance study~\cite{knapp2011} of P@$\ful$ has shown enhancement in hyperfine coupling of the phosphorous' unpaired electrons with its nucleus which is attributed to the admixture of excited states acquiring angular momentum from the cage. In contrast, the hyperfine interaction between the positively-charged muon and the unpaired electrons in the muonium atom in $\ful$, relative to free muonium, is predicted to diminish from the confinement~\cite{donzelli1996}. These discoveries render particular relevance to study open-shell atomic endofullerenes, including assessing the spectroscopy of their more stable configurations resulting from electron transfers to atomic vacancies. 

One unique phenomenon that ubiquitously occurs across endofullerene systems is the emergence of atom-fullerene ground state orbital hybridization. This entails the formation of symmetric (bonding) and antisymmetric (antibonding) hybrid states as the eigenstates of the whole system from the mixing of an atomic and a fullerene orbital of identical angular momentum symmetry. A number of our previous studies has predicted such hybrid states in various endofullerenes and their broad spectrum of photoionization properties~\cite{chakraborty2009,madjet2010,maser2012,javani2014a,javani2014b}. It is therefore of particular value to scrutinize the sensitivity of the hybridization and the photoemission dynamics of these hybrid states \textit{via} gentle changes of configuration features by, for instance, comparing the effects of confinement upon successive atoms in the periodic table. 

A prototype case that we examine here is the hybrid level photoionization of Cl@$\ful$ \textit{versus} Ar@$\ful$; Cl has just one electron less than Ar in the outer $3p$ shell and an atomic number lower by one. Closed-shell Ar, being chemically inert, almost certainly locates at the center of the spherical $\ful$. We treat the barely open-shell Cl also within the spherical geometry so as to retain the same spherical calculation as was done for the Ar case. We then consider a system of Cl$^-$@$\fulp$ produced by the transfer of a $\ful$ electron to fill in the Cl hole. This configuration attains lower energy forming closed-shell Cl$^-$. There has been experimental evidence, based on laser desorption mass spectroscopy, of $\ful$ with a single Cl$^-$ inside~\cite{zhu1994}. While it is expected that the polarization interaction of the ion can induce some offset in its position from the center of $\ful$, a density functional theory calculation with Born-Oppenheimer molecular dynamics indicates that this offset is quite small within neutral $\ful$\cite{pawar2011}. Therefore, we treat Cl$^-$@$\fulp$ assuming spherical geometry as well. We then compare the hybrid photoionization of this new configuration with Cl@$\ful$. Ultimately, the general comparison among these three endofullerene systems unfolds the delicate dependence of the hybridization and resulting photoionization cross sections on a shell-closing electron as well as on an electron transfer from the cage to the atom.
 
\section{A brief description of theory}

The details of the theory are described in Ref.\,\cite{madjet2010}. Choosing the photon polarization along the $z$-axis, the photoionization dipole transition cross section in a framework of time-dependent local density approximation (TDLDA) is given by
\begin{equation}\label{cross-pi}
\sigma_{n\ell\rightarrow k\ell'} \sim |\langle \psi_{\mathbf{k}\ell'}|z+\delta V|\phi_{n\ell}\rangle|^2.
\end{equation}
Here $\mathbf{k}$ is the momentum of the continuum electron, $z$ is the one-body dipole operator, $\phi_{nl}$ is the single electron bound wavefunction of the target level, and $\psi_{\mathbf{k}l'}$ is the respective outgoing dipole-allowed continuum wavefunction, with $l'=l\pm1$. $\delta V$ represents the complex induced potential that accounts for electron correlations within the linear response framework.

We model the bound and continuum states self-consistently using the independent particle LDA method. The jellium potentials, $V_{\mbox{\scriptsize jel}}(\mathbf{r})$, representing 60 C$^{4+}$ ions for $\ful$ is constructed by smearing the total positive charge over a spherical shell with known molecular radius $R = 3.54 \AA$~\cite{ruedel2002} and thickness $\Delta$. A constant pseudopotential $\overline{v}$ is added to the jellium for quantitative accuracy. The Kohn-Sham equations for the system of 240 electrons (four valence ($2s^22p^2$) electrons from each carbon atom), {\em plus} all electrons of the central atom with atomic number $\zeta$, are then solved in the LDA potential
\begin{equation}\label{lda-pot}
V_{\scriptsize \mbox{LDA}}(\mathbf{r}) = -\frac{\zeta}{r} + V_{\mbox{\scriptsize jel}}(\mathbf{r}) + \int d\mathbf{r}'\frac{\rho(\mathbf{r}')}{|\mathbf{r}-\mathbf{r}'|} + V_{\scriptsize \mbox{XC}}[\rho(\mathbf{r})],
\end{equation}
to obtain the bound and continuum orbitals in \eq{cross-pi}. \eq{lda-pot} uses the Leeuwen-Baerends (LB) exchange-correlation functional $V_{\scriptsize \mbox{XC}}$~\cite{van1994exchange}, which provides an accurate asymptotic description of the ground state potential. The parameters $\overline{v}$ and $\Delta$ are determined by requiring both charge neutrality and obtaining the experimental value~\cite{devries1992} of the first ionization threshold of C$_{60}$. The values of $\Delta = 1.3 \AA$ thus obtained closely agree with that extracted from measurements~\cite{ruedel2002}. We remark that treating an open-shell Cl atom in the spherical model is an approximation. For instance, our calculation overestimates Cl ionization energy from NIST data database~\cite{kramida2018} by about 7\%. But this should not take away much from the main message of this paper, particularly given that a more likely stable configuration Cl$^-$@$\fulp$ is included in this study which contains a closed-shell Cl$^-$.

The TDLDA-derived $z + \delta V(\mathbf{r})$ in \eq{cross-pi} is proportional to the induced frequency-dependent changes in the electron density~\cite{madjet2008}. This change is 
\begin{equation}\label{ind-dens}
\delta \rho (\mathbf{r}^{\prime}; \omega) = \int \chi (\mathbf{r}, \mathbf{r}^{\prime}; \omega)
z  d\mathbf{r},
\end{equation}
where the full susceptibility, $\chi$, builds the dynamical correlation from the independent-particle LDA susceptibilities
\begin{eqnarray}\label{suscep}
\chi^{0} (\mathbf{r},\mathbf{r}^{\prime };\omega) &=&\sum_{nl}^{occ}\phi _{nl}^{*}
(\mathbf{r})\phi _{nl}(\mathbf{r}^{\prime })\ G(\mathbf{r},\mathbf{r}^{\prime };\epsilon
_{nl}+\omega)  \nonumber \\
&+&\sum_{nl}^{occ}\phi _{nl}(\mathbf{r})\phi _{nl}^{*}(\mathbf{r}^{\prime })\ G^*
(\mathbf{r},\mathbf{r}^{\prime };\epsilon _{nl}-\omega)  
\end{eqnarray}
through the matrix equation $\chi = \chi^0[1-(\partial V/\partial \rho)\chi^0]^{-1}$ involving the variation of the ground-state potential $V$ with respect to the ground-state density $\rho$. The radial components of the full Green's functions in \eq{suscep} are constructed with the regular ($f_L$) and irregular ($g_L$) solutions of the homogeneous radial equation 
\begin{equation}\label{radial-eq}
\left( \frac{1}{r^2} \frac{\partial}{\partial r} r^2 \frac{\partial}{\partial r} 
     - \frac{L(L+1)}{r^2} - V_{\mbox{\scriptsize{LDA}}} 
     + E \right) f_L(g_L) (r;E) = 0
\end{equation}
as
\begin{equation}\label{green}
G_{L}(r,r^{\prime };E)=\frac{2f_{L}(r_{<};E)h_{L}(r_{>};E)}{W [f_{L},h_{L}]}  
\end{equation}
where $W$ represents the Wronskian and $h_{L}= g_{L} + i\; f_{L}$. Obviously, TDLDA thus includes the dynamical many-electron correlation by improving upon the mean-field LDA description. 
%%%%%%
\begin{figure}[h!]
\includegraphics[width=9cm]{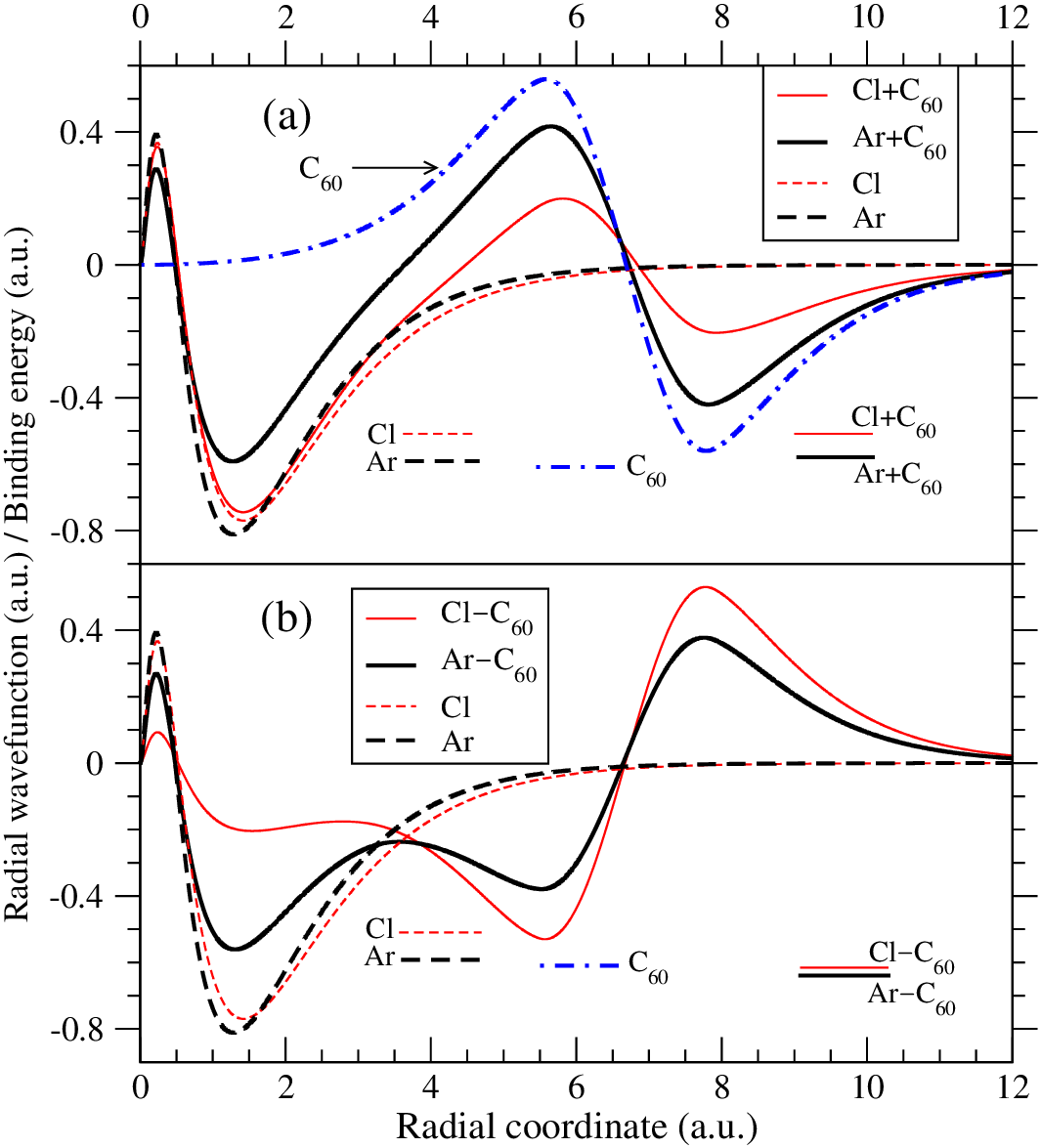}
\caption{(Color online) LDA radial symmetric (a) and antisymmetric (b) wavefunctions of Cl@$\ful$ \textit{versus} Ar@$\ful$. For both the molecules the $3p$ level of the free atom hybridizes with the $3p$ level of free $\ful$; wavefunctions of these free systems are also displayed. Relevant binding energies are also shown to aid the discussion in the text.}\label{fig1}
\end{figure}
%%%%%%%%

\section{Results and discussion}
\subsection{Cl@$\ful$ \textit{versus} Ar@$\ful$}
\subsubsection{Ground State Atom-$\ful$ Hybridization}

In an endofullerene system, an eigenstate of the free atomic hamiltonian can admix with an eigenstate of the empty $\ful$ hamiltonian of the same angular momentum symmetry to produce hybrid states which are eigenstates of the combined system. Thus, ground state LDA results for Cl@$\ful$ reveal hybridization between $3p$Cl and $3p\ful$ states which jointly produce symmetrically and antisymmetrically combined states of Cl@$\ful$ that can be written as,
\begin{subequations}\label{bound-hyb}
\begin{equation}\label{bound-hib1}
|\mbox{Cl}+\mbox{C}_{60}\rangle = |\phi_+\rangle = \sqrt{\alpha}|\phi_{3p \mbox{\scriptsize Cl}}\rangle + \sqrt{1-\alpha}|\phi_{3p \fuls}\rangle
\end{equation}
\begin{equation}\label{bound-hib2}
|\mbox{Cl}-\mbox{C}_{60}\rangle = |\phi_-\rangle = \sqrt{1-\alpha}|\phi_{3p \mbox{\scriptsize Cl}}\rangle - \sqrt{\alpha}|\phi_{3p \fuls}\rangle
\end{equation}
\end{subequations}
where the fraction $\alpha$ is the mixing parameter that renders the states orthonormal. In Fig.\,1, the radial components of these wavefunctions are shown and compared with the corresponding hybrid wavefunctions of Ar@$\ful$. From a perturbation theory viewpoint, the strength of this mixing is proportional directly to the overlap of the participating (free) orbitals and inversely to the separation of their binding energies. As Fig.\,1 indicates, the energy of $3p$Ar is extremely close to that of $3p\ful$, while $3p$Cl moves a bit higher, leading to a stronger mixing with a value of $\alpha$ (\eq{bound-hyb}) close to about 0.5 (equal share of atom-fullerene character) for Ar@$\ful$. A somewhat reduced hybridization in Cl@$\ful$ with a greater value of $\alpha$ thus implies enhanced Cl and enhanced $\ful$ characters, respectively, for the symmetric and antisymmetric state. This occurs in spite of a slightly increased wavefunction overlap due to a small displacement of $3p$Cl wavefunction toward the shell from that of $3p$Ar (Fig.\,1). Also note that the resulting symmetric hybrids of the systems are more separated energetically than the antisymmetric hybrids. 
%%%%%%
\begin{figure}[h!]
\includegraphics[width=9cm]{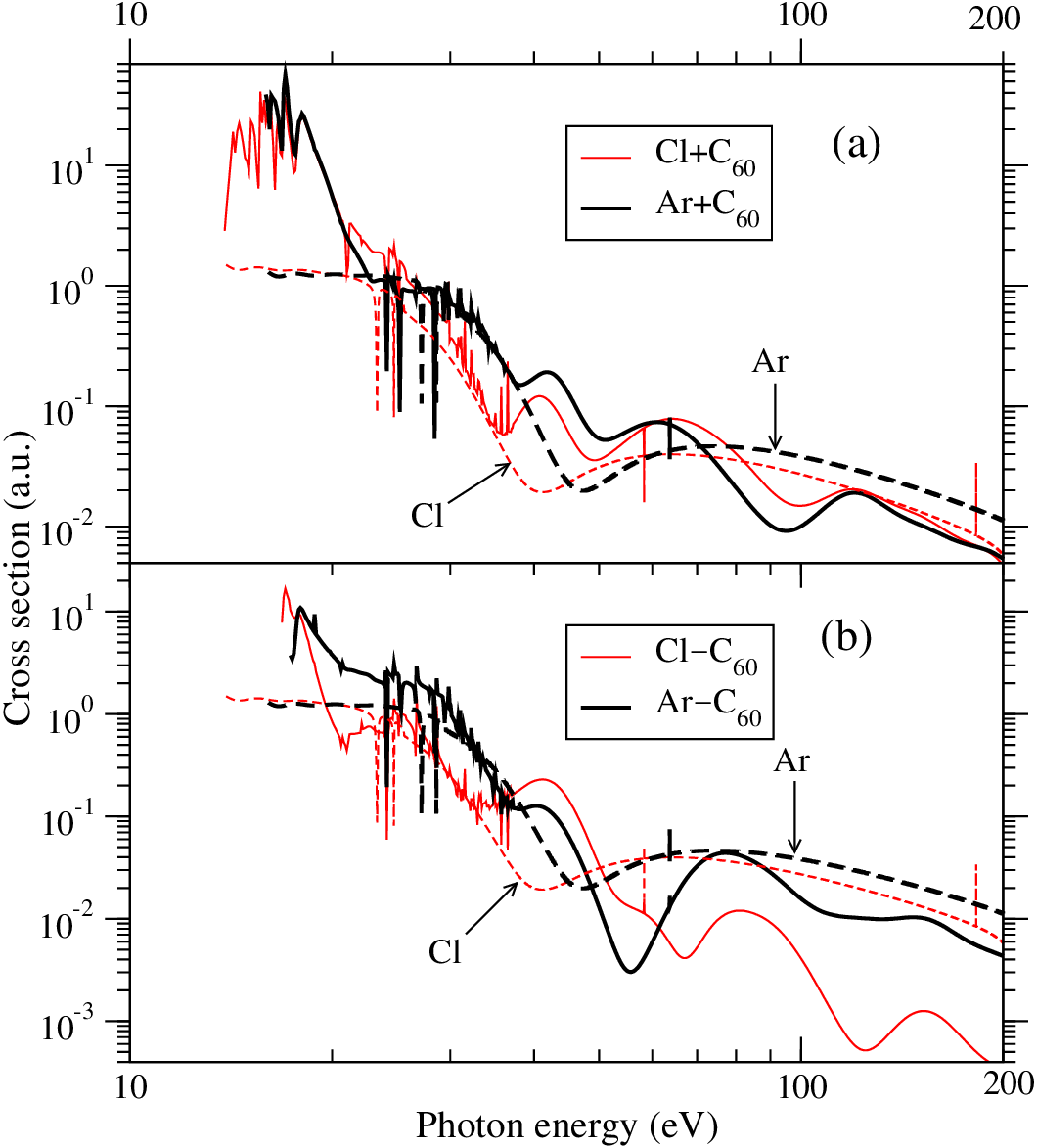}
\caption{(Color online) TDLDA photoionization cross sections of symmetric (a) and antisymmetric (b) levels of Cl@$\ful$ compared to Ar@$\ful$. Cross sections for $3p$ of free Cl and free Ar are also plotted for comparisons.} \label{fig2}
\end{figure}
%%%%%%

\subsubsection{Photoionization of Hybrid Levels}

Cross sections calculated at the correlated TDLDA level for the hybrid states of Cl@$\ful$ and Ar@$\ful$ are presented in Fig.\,2 as a function of the photon energy. Comparing these with $3p$ of free Cl and Ar indicates plasmon driven enhancements at low energies~\cite{madjet2007,javani2014a}. This enhancement is significantly stronger for symmetric photoemission than the antisymmetric one. In the framework of interchannel coupling due to Fano, the correlation-modified (TDLDA) matrix element of the photoionization of X$\pm\ful$, X being Cl or Ar, can be written as \cite{javani2014a},
\begin{eqnarray}\label{gen-mat-element}
{\cal M}_\pm (E) &=& {\cal D}_\pm (E)\nonumber \\
                 &+& \!\!\!\!\displaystyle\sum_{n\ell}\!\!\!\int\!\!\! dE' \!\frac{\langle\psi_{n\ell}(E')|\frac{1}{|{\bf r}_{\pm}-{\bf r}_{n\ell}|}
|\psi_{\pm}(E)\rangle}{E-E'} \!{\cal D}_{n\ell} (E')
\end{eqnarray}
in which the single electron (LDA) matrix element is
\begin{equation}\label{se-mat-element}
{\cal D}_\pm (E) = \langle ks(d)|z|\phi_\pm\rangle
\end{equation}
and $|\psi_{nl}\rangle$ in the interchannel coupling integral is the (continuum) wavefunction of the $n\ell\rightarrow k\ell'$ channel. Taking the hybridization into account, the channel wavefunctions in \eq{gen-mat-element} become
\begin{subequations}\label{channel-hyb}
\begin{equation}\label{channel-hib1}
|\psi_+\rangle = \sqrt{\alpha}|\psi_{3p \mbox{\scriptsize X}}\rangle + \sqrt{1-\alpha}|\psi_{3p \fuls}\rangle
\end{equation}
\begin{equation}\label{channel-hib2}
|\psi_-\rangle = \sqrt{1-\alpha}|\psi_{3p \mbox{\scriptsize X}}\rangle - \sqrt{\alpha}|\psi_{3p \fuls}\rangle.
\end{equation}
\end{subequations}  

Substituting Eqs.\ (\ref{bound-hyb}), but for a general X, and (\ref{channel-hyb}) in \eq{gen-mat-element}, and noting that the overlap between a pure X and a pure $\ful$ bound state is negligible, we separate the atomic and fullerene contributions to the integral to get the TDLDA matrix element for X$\pm\ful$ levels as,
\begin{subequations}\label{gen-mat-element+-}
\begin{equation}\label{gen-mat-element+}
{\cal M}_+ (E) = \sqrt{\alpha}{\cal M}_{3p \mbox{\scriptsize X}} (E) + \sqrt{1-\alpha}{\cal M}_{3p \fuls} (E)
\end{equation}
\begin{equation}\label{gen-mat-element-}
{\cal M}_- (E) = \sqrt{1-\alpha}{\cal M}_{3p \mbox{\scriptsize X}} (E) - \sqrt{\alpha}{\cal M}_{3p \fuls} (E),
\end{equation} 
\end{subequations}
where the second terms on the right hand side are responsible for the plasmonic enhancements at the lower energies, as seen in Fig.\,2. Writing the atomic and $\ful$ contributions in \eq{gen-mat-element+-} respectively as complex quantities $X_r+iX_i$ and $C_r+iC_i$, and recalling that the cross section $\sigma$ is proportional to the square modulus of the matrix element, we can express the cross sections as,
\begin{subequations}\label{cross}
\begin{equation}\label{cross+}
\sigma_+(E) \sim (\sqrt{\alpha}X_r+\sqrt{1-\alpha}C_r)^2 + (\sqrt{\alpha}X_i+\sqrt{1-\alpha}C_i)^2
\end{equation}
\begin{equation}\label{cross-}
\sigma_-(E) \sim (\sqrt{1-\alpha}X_r-\sqrt{\alpha}C_r)^2 + (\sqrt{1-\alpha}X_i-\sqrt{\alpha}C_i)^2. 
\end{equation}
\end{subequations}
Evidently, the enhancement of the symmetric state cross section, \eq{cross+}, involving the sum of real and imaginary components, will be universally larger than that of the antisymmetric state which include their differences. Indeed, this is seen for both Cl@$\ful$ and Ar@$\ful$ in Fig.\,2 and a direct consequence of the in-phase \textit{versus} out-of-phase radial oscillation of the hybrid wavefunctions at $\ful$ shell (Fig.\,1). However, the detailed similarities and differences of enhanced cross section between two systems must depend on their respective values of $\alpha$ in a rather complicated way that involves the interferences between the atomic and the $\ful$ components of the matrix elements in \eq{cross}. We note that while this enhancement is of a comparable size for the symmetric states of both systems in Fig.\,2(a), it is significantly weaker for antisymmetric Cl-$\ful$ compared to Ar-$\ful$ [Fig\,2(b)]. The situation further complicates from the effect of lowering binding energies of Cl@$\ful$ hybrid levels as is discussed in subsection IIIB below.

As the plasmonic effect weakens with increasing energy, the cross sections largely follow their free atom curves, as seen in Fig.\,2. Cooper minimum-like structures develop~\cite{dixit2013} around 50 eV on both the symmetric curves with the minimum of Cl@$\ful$ being deeper due to its higher atomic character. Note that the Cooper minima in $3p$ cross sections for free atoms are clearly visible in Fig.\,2. In any case, such minima also show up around 60 eV for antisymmetric emissions where the structure for Cl@$\ful$ is very weak because of its weaker atomic character.    

Above these energies the cross sections oscillate as a consequence of a well-known multipath interference mechanism~\cite{cdm2000} due to the cavity structure of $\ful$ which was modeled earlier~\cite{mccune2009}. At such high energies the interchannel coupling in \eq{gen-mat-element} vanishes to simplify Eqs.\,(\ref{gen-mat-element+-}) to,
\begin{subequations}\label{ip-mat-element+-}
\begin{equation}\label{gen-mat-element+}
{\cal D}_+ (E) = \sqrt{\alpha}{\cal D}_{3p \mbox{\scriptsize X}} (E) + \sqrt{1-\alpha}{\cal D}_{3p \fuls} (E)
\end{equation}
\begin{equation}\label{gen-mat-element-}
{\cal D}_- (E) = \sqrt{1-\alpha}{\cal D}_{3p \mbox{\scriptsize X}} (E) - \sqrt{\alpha}{\cal D}_{3p \fuls} (E).
\end{equation} 
\end{subequations}
The multipath interference model gives~\cite{mccune2009}
\begin{subequations}\label{etas-amp}
\begin{eqnarray}\label{etas@-amp-atom}
{\cal D}_{3p \mbox{\scriptsize X}} &\sim& {\cal D}^{\mbox{\tiny atom}}(k) \nonumber \\
                  & + & A^{\mbox{\tiny refl}}(k)\left[e^{-ikD_o}e^{-iV_0\frac{2\Delta}{k}} - e^{-ikD_i} \!\right]
\end{eqnarray}
\begin{equation}\label{etas@-amp-shell}
{\cal D}_{3p \fuls} \sim  A^{\mbox{\tiny shell}}(k)e^{-i\frac{V_0}{k}}\left[a_ie^{-ikR_i}- a_oe^{-ikR_o}\!\right],
\end{equation}
\end{subequations}
where the photoelectron momentum $k =\sqrt{2(E-\epsilon_{\pm})}$ in atomic units, $a_i$ and $a_o$ are the values of $\phi_{\pm}$ at the inner and outer radii $R_i$ and $R_o$ of $\ful$, and $V_0$ is the average depth of the shell potential. In \eq{etas@-amp-atom}, while ${\cal D}^{\mbox{\tiny atom}}$ is the contribution from the atomic region, the second term denotes the reflection induced oscillations in momentum coordinates with amplitude $A^{\mbox{\tiny refl}}$ and frequencies related to $D_i$ and $D_o$, the inner and outer diameters of the shell. Since, obviously, $A^{\mbox{\tiny refl}}$ is proportional to ${\cal D}^{\mbox{\tiny atom}}$, the larger the atomic component of a hybrid wavefunction, the stronger is the reflection and the higher is the chances that the oscillations occur about the free atom result. This is exactly what is seen for the high energy cross section of Cl+$\ful$ in Fig.\,2(a). On the other hand, \eq{etas@-amp-shell} presents the portion of the overlap integral from the shell region, producing two collateral emissions from shell edges, which oscillate in frequencies related to $R_i$ and $R_o$. This part will dominate if a hybrid level has a stronger $\ful$ character, like for Cl-$\ful$, which intensifies oscillations at higher energies but falls significantly lower than free $3p$Cl [Fig.\,2(b)]. For the Ar@$\ful$ hybrids, however, due to their almost equal share of atom-$\ful$ character (Fig.\,1), the strength of high-energy cross sections are comparable, somewhat below $3p$Ar, but the differences in the details of their shapes again owe to the interference between reflective and collateral emissions.
%%%%%%
\begin{figure}[h!]
\includegraphics[width=9cm]{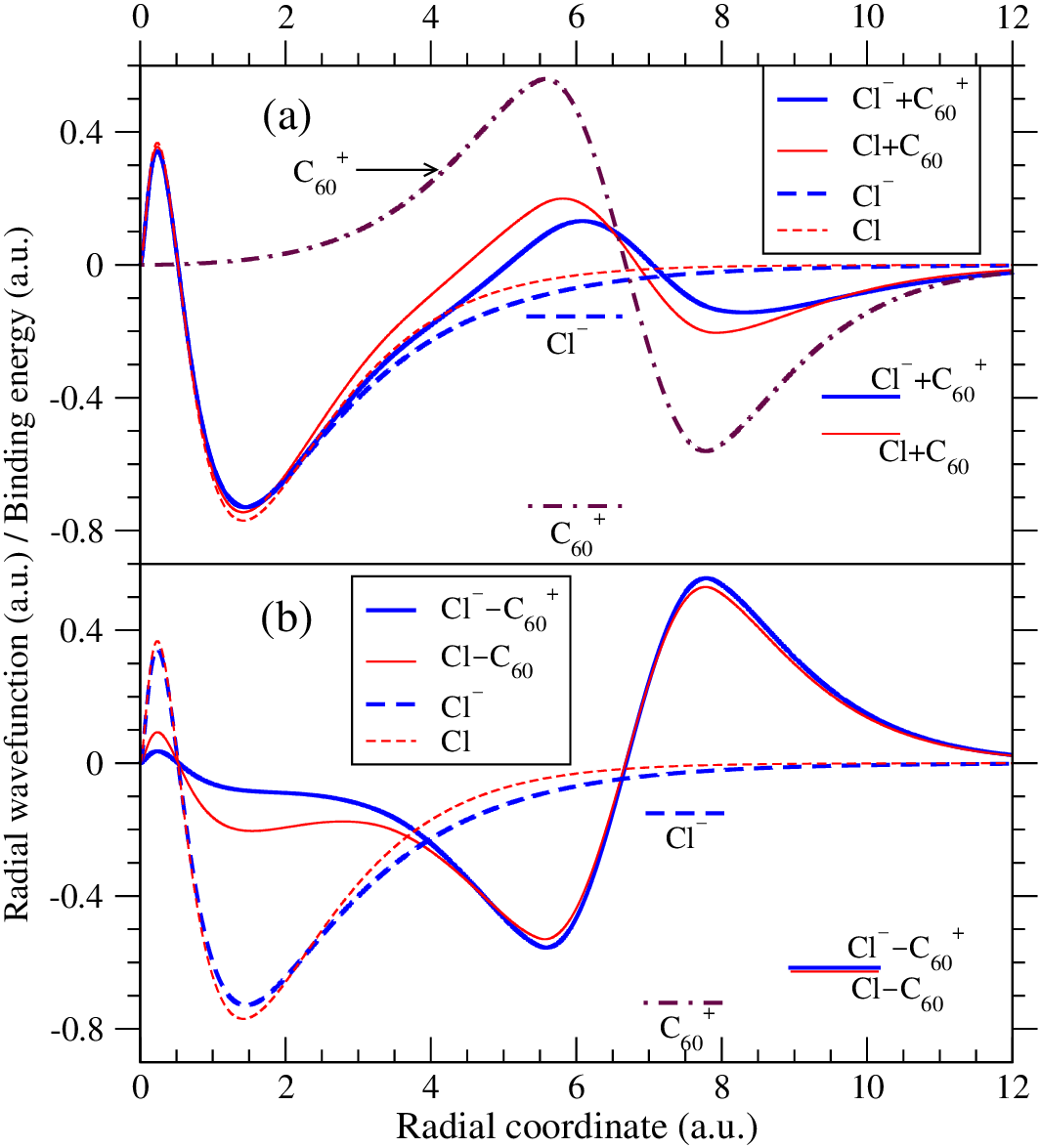}
\caption{(Color online) Same as Fig.\,1 but for Cl@$\ful$ in comparison with a more stable configuration Cl$^-$@$\fulp$ of the molecule. Participating wavefunctions of the free systems Cl, Cl$^-$ and $\fulp$, and relevant binding energies are included.} \label{fig4}
\end{figure}
%%%%%%%%
%%%%%%
\begin{figure}[h!]
\includegraphics[width=9cm]{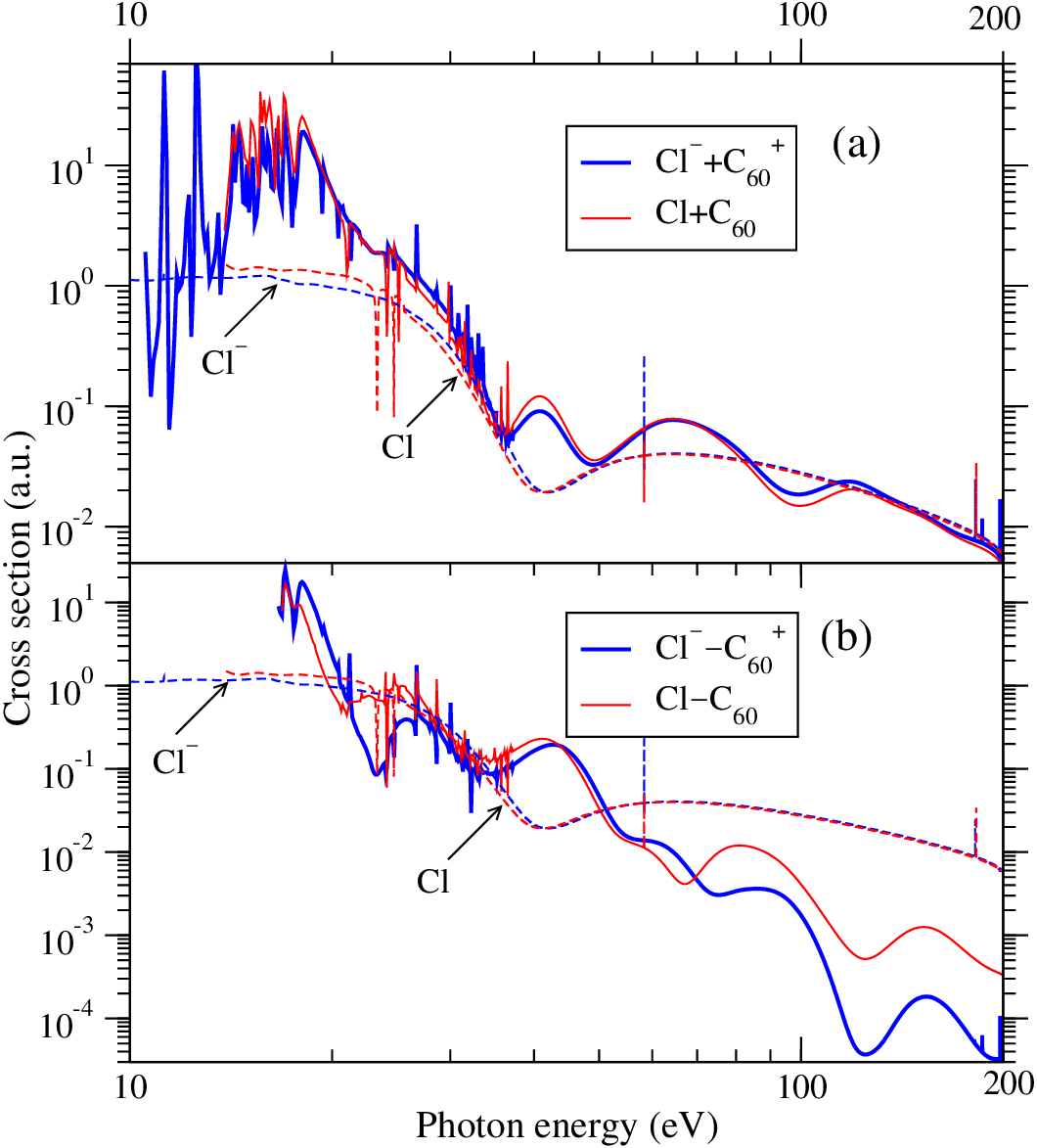}
\caption{(Color online) Same as Fig.\,2 but for Cl@$\ful$ \textit{versus} Cl$^-$@$\fulp$ configuration. TDLDA cross sections for $3p$ of free Cl and free Cl$^-$ are also shown.} \label{fig4}
\end{figure}
%%%%%%%%

\subsection{Cl@C$_{60}$ \textit{versus} Cl$^-$@$\fulp$}

A reactive Cl atom is very likely to capture an electron from $\ful$ which will bring the compound to a more stable configuration Cl$^-$@$\fulp$. For an empty $\fulp$, our LDA ground state structure is insensitive to the location of the hole among the molecular levels which is not too surprising for a cloud of 240 delocalized electrons. Likewise, for the complex Cl$^-$@$\fulp$ the LDA energies and wavefunctions of all pure fullerene states are found independent of which $\ful$ orbital the hole is situated at. In fact, these energies and wavefunctions remain practically identical to those of empty $\ful$, and therefore of little interest. However, the hybridization between $3p$Cl$^-$ and $3p$$\fulp$ is somewhat modified from that in Cl@$\ful$, or, in other words, between after and before the electron makes a transition, as we describe below. Yet it is found that the $3p$ hybridization in Cl$^-$@$\fulp$ is still insensitive to the $\ful$ level the electron transitions from. Therefore, our results presented here are robust being free of a specific choice of the hole level. On the other hand, since the Cl$^-$ ion has a higher binding energy than the neutral Cl, the stable configuration Cl$^-$@$\fulp$ exhibits a lower total energy than Cl@$\ful$, and therefore should be more abundantly formed. Also, Cl$^-$, being a closed-shell system, will be more accurately described by our spherical LDA model than open-shell Cl. 

The $3p$ hybridization in Cl$^-$@$\fulp$ is compared to that in Cl@$\ful$ in Fig.\,3. Note that for Cl$^-$@$\fulp$ the participating free levels $3p$Cl$^-$ and $3p$@$\fulp$ are significantly apart from each other (Fig.\,3), the former becoming quite shallower and the latter becoming deeper compared to their counterparts in Cl@$\ful$ (Fig.\,1). This strongly disfavors the hybridization. Conversely, as also seen in Fig.\,3, the radial wavefunction of $3p$Cl$^-$ extends more radially outward than $3p$Cl to increase its overlap with $3p$$\fulp$ which favors the hybridization. In the tug-of-war between these two effects, the former wins resulting in some modification of hybridization for Cl$^-$@$\fulp$ configuration as displayed in Fig.\,3. It is seen that the atomic character stays unchanged, but the $\ful$ character weakens at the shell for the symmetric hybrid, while the reverse is true for the antisymmetric hybrid. Comparing the energy of the hybrid levels, the symmetric hybrid moves energetically higher [Fig.\,3(a)] while the antisymmetric hybrids barely separate [Fig.\,3(b)] as a result of the charge transfer. And this modified hybridization affects the resulting photoionization cross sections.

Fig.\,4 compares the TDLDA cross sections for photoionizations from hybrid levels of Cl$^-$@$\fulp$ with Cl@$\ful$. Note that even though there is a slight modification in hybridization (Fig.\,3) post electron transfer, the cross section of the symmetric level [Fig.\,4(a)] is hardly modified, except that Cl$^-$+$\fulp$ starts at a lower photon energy due to the reduction of its binding energy. To be more precise, some reduction in the intensity of the structure of Cl$^-$+$\fulp$ wavefunction [Fig.\,3(a)] in the shell region results in little effect on the cross section. In the energy region of the plasmon, this can be understood in general from the interchannel coupling contribution of the matrix element in \eq{gen-mat-element}. Specifically, this term in part embodies the overlaps of the ``fractional" ionization channel emanating from the shell region of the symmetric hybrid with all $\ful$ channels which are accounted for in ${\cal M}_{3p \fuls}$ in \eq{cross+}. The channel overlap includes both the overlap between the bound and the continuum wavefunctions. While the reduction of the structure in Cl$^-$+$\fulp$ radial wave noted above lowers the aggregated strength of bound overlaps, the continuum overlaps will be slightly favored due to the following reason. A continuum overlap is more efficient at higher photoelectron momenta $k$, since differences between the momenta due to different level energies reduce enabling the continuum waves to oscillate progressively in phase with each other. Therefore, the Cl$^-$+$\fulp$ state opening at a lower photon energy benefits continuum overlaps to cause its net increase. This gain must be compensating the loss due to weakening hybridization to affect practically no change in the symmetric hybrid level cross section [Fig.\,4(a)] after the electron transfers. Note that the term ${\cal M}_{3p \mbox{\scriptsize X}}$ in \eq{cross+} has no effect here due to about the same atomic character of hybrid wavefunctions before and after the electron switches locations. For the higher energy emission of this hybrid [Fig.\,4(a)], a rather miniscule weakening of the multipath oscillations following the electron transfer traces to the reduction of its $\ful$ character that slightly reduces the collateral emission in \eq{etas-amp}, while keeping the reflective emission unchanged. The strong atomic character of this hybrid, which remains unchanged in spite of the electron transition, keeps the average strength of both cross sections close to $3p$ of Cl and Cl$^-$ which are practically equal at these energies. 

Differences between the cross sections of asymmetric hybrid states of Cl$^-$@$\fulp$ and Cl@$\ful$, shown in Fig.\,4(b), are rather strong. The differences at plasmonic energies up to 40 eV are due to the reduction of the atomic character [Fig.\,3(b)], particularly \textit{via} the interference effects illustrated in \eq{cross-}. Note that since the binding energies of this hybrid state suffer a very little change upon the charge transfer, the continuum overlap effect described above does not apply. However, at higher energies, 80 eV and above, the difference in cross sections grows progressively stronger. This must be due a cumulative effect of the reduction of the atomic component of direct ionization, as a result of the reduced atomic character in Cl$^-$-$\fulp$, the subsequent reduction of reflective amplitude in \eq{etas-amp}, and the interference between them. To this end, even after the molecule evolves to a stable configuration Cl$^-$@$\fulp$, the strong plasmonic magnification of the emission from the symmetric hybrid remains, but the high energy response of the asymmetric state substantially changes. 

\section{Conclusions}      

Using a fairly successful methodology of the time-dependent local density approximation based on a spherical jellium modeling of $\ful$'s ion core, we compute the photoemission cross sections of the atom-fullerene hybrid levels of Cl@$\ful$. A comparison of the results with those of Ar@$\ful$ probes the modification effects of a shell-closing electron on the properties of these hybrid photoemissions. However, Cl@$\ful$ must be an unstable system and will likely induce an electron transfer from $\ful$ to Cl to reach a more stable and lower energy configuration. We compared the results between these two configurations of Cl endofullerenes to assess the effects of this electron transition on the hybrid photodynamics. Tuning the configuration gently along the sequence of Ar@$\ful$ to Cl@$\ful$ to Cl$^-$@$\fulp$, a systematic evolution of the ionizing response properties of hybrid states is uncovered. This is the first study of the photoionization properties of a halogen endofullerene to the best of our knowledge. A strong magnification of the low energy emission of the symmetric hybrid over the entire giant plasmon resonance region and an enhanced multipath interference effects for the high energy antisymmetric hybrid are found to be the most robust features in our study.  

\begin{acknowledgments}
The research is supported by the US National Science Foundation Grant No.\ PHY-1806206 (HSC) and the US Department of Energy, Office of Science, Basic Energy Sciences, under award DE-FG02-03ER15428 (STM). 
\end{acknowledgments}


\begin{thebibliography}{10}

\bibitem{popov2013} 
A.A. Popov, S. Yang, and L. Dunsch, Endohedral fullerenes, Chem.\ Rev. \textbf{113}, 5989 (2013).

\bibitem{mueller2008} 
A. M\"{u}ller, S. Schippers, M. Habibi, D. Esteves, J.C. Wang, R.A. Phaneuf, A.L.D. Kilcoyne, A. Aguilar, and L. Dunsch, Significant redistribution of Ce 4$d$
oscillator strength observed in photoionization of endohedral Ce@C$_{82}^+$ ions, \prl\, \textbf{101}, 133001 (2008).

\bibitem{kilcoyne2010}
A.L.D. Kilcoyne, A. Aguilar, A. M\"{u}ller, S. Schippers, C. Cisneros, G. Alna'Washi, N.B. Aryal, K.K. Baral, D.A. Esteves, C.M. Thomas, and R.A. Phaneuf, Confinement resonances in photoionization of Xe@$\fulp$, \prl\, \textbf{105}, 213001 (2010).

\bibitem{phaneuf2013}
R.A. Phaneuf, A.L.D. Kilcoyne, N.B. Aryal, K.K. Baral, D.A. Esteves-Macaluso, C.M. Thomas, J. Hellhund, R. Lomsadze, T.W. Gorczyca, C.P. Balance,  S.T. Manson, M.F. Hasoglu, S. Schippers, and A. M\"{u}ller, Probing confinement resonances by photoionizing Xe inside a $\fulp$ molecular cage, \pra\, \textbf{88}, 053402 (2013).

\bibitem{dolmatove2009}
V.K. Dolmatov, Photoionization of atoms encaged in spherical fullerenes, \textit{Theory of Confined Quantum Systems: Part two, Advances in Quantum Chemistry}, edited by J.R. Sabin and E. Braendas (Academic Press, New York, 2009), Vol. 58, pp. 13-68.

\bibitem{chakraborty2015}
H.S. Chakraborty and M. Magrakvelidze, Many-electron response of gas-phase fullerene materials to ultraviolet and soft X-ray photons, \textit{From Atomic to Mesoscale: the Role of Quantum Coherence in Systems of Various Complexities}, edited by S. Malinovskaya and I. Novikova (World Scientific, Singapore, 2015) p. 221.

\bibitem{harneit2007}
W. Harneit, C. Boehme, S. Schaefer, K. Huebner, K. Fortiropoulos, and K. Lips, Room temperature electronic detection of spin coherence in $\ful$, \prl\, \textbf{98}, 216601 (2007).

\bibitem{ju2011}
C. Ju, D. Suter, and J. Du, An endohedral fullerene-based nuclear spin quantum computer, Phys.\ Lett.\ A  375, \textbf{1441} (2011).

\bibitem{takeda2006}
A. Takeda, Y. Yokoyama, S. Ito, T. Miyazaki, H. Shimotani, K. Yakigaya, T. Kakiuchi, H. Sawa, H. Takagi, K. Kitazawa, and N. Dragoe, Superconductivity of doped Ar@$\ful$, Chem.\ Commun. \textbf{8}, 912 (2006).

\bibitem{melanko2009} 
J.B. Melanko, M.E. Pearce, and A.K. Salem, \textit{Nanotechnology in Drug Delivery}, edited by M.M. de Villiers, P. Aramwit, and G.S. Kwon (Springer, New York, 2009) 105.

\bibitem{ross2009} 
R.B. Ross, C.M. Cardona, D.M. Guldi, S.G. Sankaranarayanan, M.O. Reese, N. Kopidakis, J. Peet, B. Walker, G.C. Bazan, E.V. Keuren, B.C. Holloway, and M. Drees, Endohedral fullerenes for organic photovoltaic devices, Nature Materials \textbf{8}, 208 (2009).
 
\bibitem{becker2000} 
L. Becker, R.J. Poreda, and T.E. Bunch, Fullerenes: An extraterrestrial carbon carrier phase for noble gases,  Proc.\ Nat.\ Acad.\ Sc.\ of USA \textbf{97}, 2979 (2000).

\bibitem{lawler2017} 
R.G. Lawler, Nonmetallic Endofullerenes and the Endohedral Environment: Structure, Dynamics, and Spin Chemistry, \textit{Endohedral Fullerenes: Electron Transfer and Spin} edited by A. Popov (Springer, 2017) p. 229.

\bibitem{morton2007} J. J. L. Morton, A. M. Tyryshkin, A. Ardavan, K. Porfyrakis, S.A. Lyon, and G. A. D. Briggs, Environmental effects on electron spin relaxation in N@$\ful$, \prb\, \textbf{76}, 085418 (2007).

\bibitem{knapp2011} 
C. Knapp, N. Weiden, H. Kass, K.-P. Dinse, B. Pietzak, M. Waiblinger, and A. Weidinger, Electron paramagnetic resonance study of atomic phosphorus encapsulated in [60]fullerene, Molecular Physics \textbf{95}, 999 (1998); online (2011).

\bibitem{donzelli1996} 
O. Donzelli, T. Briere, and T. P. Das, Location of muonium and hydrogen in C60 fullerene and associated electronic structure and hyperfine properties, Hyperfine Interactions \textbf{97}, 19 (1996).

\bibitem{chakraborty2009} 
H.S. Chakraborty, M.E. Madjet, T. Renger, Jan-M. Rost, and S.T. Manson, Photoionization of hybrid states in endohedral fullerenes, \pra\, \textbf{79}, 061201(R) (2009).

\bibitem{madjet2010} 
M.E. Madjet, T. Renger, D.E. Hopper, M.A. McCune, H.S. Chakraborty, Photoionization of Xe inside $\ful$: Atom-fullerene hybridization, giant cross-section enhancement, and correlation confinement resonances, Jan-M Rost, and S.T. Manson, \pra\, \textbf{81}, 013202 (2010).

\bibitem{maser2012} 
J.N. Maser, M.H. Javani, R. De, M.E. Madjet, H.S. Chakraborty, and S.T. Manson, Atom-fullerene hybrid photoionization mediated by coupled $d$ states in Zn@$\ful$, \pra\, \textbf{86}, 053201 (2012).

\bibitem{javani2014a} 
M.H. Javani, R. De, M.E. Madjet, S.T. Manson, and H.S. Chakraborty, Photoionization of bonding and antibonding-type atom-fullerene hybrid states in Cd@$\ful$ vs  Zn@$\ful$, J.\ Phys.\ B \textbf{47}, 175102 (2014).

\bibitem{javani2014b} 
M.H. Javani, J.B. Wise, R. De, M.E. Madjet, S.T. Manson, and H.S. Chakraborty, Resonant Auger-intercoulombic hybridized decay in the photoionization of endohedral fullerenes, \pra\, \textbf{89}, 063420 (2014). 

\bibitem{zhu1994} 
L. Zhu, S. Wang, Y. Li, Z. Zhang, H. Hou, and Q. Qin, Evidence for fullerene with single chlorine anion inside, \apl\, \textbf{66}, 702 (1994).

\bibitem{pawar2011} 
P. Ravinder and V. Subramanian, Studies on the encapsulation of various anions in different fullerenes using density functional theory calculations and Born-Oppenheimer molecular dynamics simulation, J.\ Phys.\ Chem.\ A \textbf{115}, 11723 (2011).

\bibitem{ruedel2002} 
A. R\"{u}del, R. Hentges, H.S. Chakraborty, M.E. Madjet, and J.M. Rost, Imaging delocalized electron clouds: Photoionization of $\ful$ in Fourier reciprocal space, \prl\, \textbf{89}, 125503 (2002).

\bibitem{van1994exchange} 
R.~Van~Leeuwen and E.~J. Baerends, Exchange-correlation potential with correct asymptotic behavior, \pra\, {\bf 49}, 2421 (1994).

\bibitem{devries1992} 
J. de Vries, H. Steger, B. Kamke, C. Menzel, B. Weisser, W. Kamke, I.V. Hertel, Single-photon ionization of $\ful$- and C$_{70}$-fullerene with synchrotron radiation: determination of the ionization potential of $\ful$, Chem.\ Phys.\ Lett. {\bf 188}, 159 (1992).

\bibitem{kramida2018}
A. Kramida, Yu. Ralchenko, J. Reader, and NIST ASD Team, NIST Atomic Spectra Database (ver. 5.6.1), (2018). 

\bibitem{madjet2008} 
M.E. Madjet, H.S. Chakraborty, J.M. Rost, and S.T. Manson, Photoionization of $\ful$: a model study, J.\ Phys.\ B \textbf{41}, 105101 (2008).

\bibitem{madjet2007} M.E. Madjet, H.S. Chakraborty, and S.T. Manson, Giant enhancement in low energy photoemission of Ar confined in $\ful$, \prl\, \textbf{99}, 243003 (2007).

\bibitem{dixit2013} 
G. Dixit, H.S. Chakraborty, and M.E. Madjet, Time delay in the recoiling valence photoemission of Ar endohedrally confined in $\ful$, \prl\, {\bf 111}, 203003 (2013).

\bibitem{cdm2000} J.-P. Connerade, V.K. Dolmatov, and S.T. Manson, On the nature and origin of confinement resonances, J.\ Phys.\ B \textbf{33}, 2279 (2000).

\bibitem{mccune2009} M.A. McCune, M.E. Madjet, H.S. Chakraborty, Reflective and collateral photoionization of an atom inside a fullerene: Confinement geometry from reciprocal spectra, \pra\, \textbf{80}, 011201(R) (2009).

\end{thebibliography}
\end{document}